\documentclass[12pt,a4paper,final,onecolumn]{IEEEtran}

\usepackage{upgreek}
\usepackage{booktabs}
\usepackage{cite}

\ifCLASSINFOpdf
\usepackage[pdftex]{graphicx}
\else
\fi

\usepackage{amsmath}
  \usepackage[caption=false,font=footnotesize]{subfig}

\begin{document}
\IEEEoverridecommandlockouts


\title{A new model for the TCAD simulation of the silicon damage by high fluence proton irradiation}

\author{{Joern~Schwandt,
        Eckhart~Fretwurst,
        Erika~Garutti,
        Robert~Klanner, Christian Scharf, and Georg Steinbrueck}
\thanks{J.~Schwandt, E. Fretwurst, E.~Garutti, R.~Klanner and G.~Steinbrueck  are with the Institute of Experimental Physics, University of Hamburg, Luruper Chaussee 149, D-22761 Hamburg, Germany.}
\thanks{
C.~Scharf now at Institut fuer Physik, Humboldt University zu Berlin, Newtonstr. 15, D-12489 Berlin, Germany.}
 \thanks{J.~Schwandt, the corresponding author, can be contacted by \emph{Email: \mbox{ {joern.schwandt@desy.de}}} and \emph{Tel.: +49~40~8998~4742}. } }

\maketitle

\begin{abstract}
For the high-luminosity phase of the Large Hadron Collider (HL-LHC), at the 
expected position of the innermost pixel detector layer of the CMS and ATLAS 
experiments, the estimated equivalent neutron fluence after 3000~fb$^{-1}$ 
is 2$\cdot$10$^{16}$~n$_{eq}$/cm$^2$, and the IEL (Ionizing Energy Loss) dose
in the SiO$_2$ 12~MGy. The optimisation of the pixel sensors and the 
understanding of their performance as a function of fluence and dose makes a 
radiation damage model for TCAD simulations, which describes the available 
experimental data, highly desirable. The currently available bulk-damage 
models are not able to describe simultaneously the measurements of dark current~(I-V),
capacitance-voltage~(C-V) and charge collection efficiency~(CCE) of pad diodes for
fluences $\ge 1\cdot 10^{15}$~n$_{eq}$/cm$^2$. 
Therefore, for the development and validation of a new accurate bulk damage model 
we use I-V, C-V and CCE measurements on pad diodes available within 
the CMS-HPK campaign and data from samples irradiated recently with 24~GeV/c protons. 
For the determination of the radiation-induced damage parameters we utilise the 
"optimiser" of Synopsys TCAD, which allows the minimisation of the difference
between the measured and simulated I-V, C-V and CCE. The outcome of this optimisation,
the Hamburg Penta Trap Model (HPTM), provides a consistent and accurate description of
the measurements of diodes irradiated with protons in the fluence range from 
3$\cdot$10$^{14}$~n$_{eq}$/cm$^2$ to 1.3$\cdot$10$^{16}$~n$_{eq}$/cm$^2$.

\end{abstract}


%

\section{Introduction}
%
%
%
%
\IEEEPARstart{A}{ttempts} to understand and optimise 
irradiated silicon sensors using TCAD simulations have been around 
for quite some time \cite{Chiochia:2006,Petasecca:2006,Pennicard:2008,
Peltola:2015,Perugia:2015}.
The biggest challenge for such device simulations is the implementation of a
satisfactory model for radiation damage effects. In the attempt to implement a model 
based on defect spectroscopy, one has to consider that for example, hadron irradiation
introduces more than 10~different types of electrically active point 
and cluster defects in the silicon band-gap. For fluences 
$\ge 10^{15}$~n$_{eq}$/cm$^2$, microscopic measurements 
like Thermally Stimulated Current techniques 
are not possible and for lower fluences only partial information 
about the defects is available. Furthermore, it is currently not possible to simulate 
cluster defects in TCAD \cite{Donegani:2018gw}.
Therefore, "effective models" are developed which assume a minimum number of point defect 
levels, and tune the parameters to reproduce macroscopic measurements. 

\section{Optimisation}
For our study we used 200 $\mu$m thick float zone p-type diodes with an area 
of 5$\times$5~mm$^2$ from the CMS-HPK campaign. 
The diodes were irradiated at the CERN PS with 24~GeV/c 
protons to fluences of 0.3, 1, 3, 6, 8 and $13\cdot 10^{15}$~n$_{eq}$/cm$^2$ and 
annealed at 60$^\circ$C
for 80~min. I-V, C-V and CCE measurements with infrared light have been performed at 
$-20^\circ$C and $-30^\circ$C. All simulations have been performed with 
Synopsys TCAD \cite{synopsys}.
For the simulation of the I-V and C-V dependence a 1D~model 
is used. The doping profiles of the $n^+$ and $p^+$ implants are taken from spreading
resistance measurements. The bulk doping and the active thickness were adjusted to 
match the C-V measurements of the non-irradiated diode. As with I-V and C-V 
measurements alone one cannot obtain a model which describes the CCE from infrared light
we used in the optimisation a simulation of the CCE from infrared light at 
5~voltage steps. For this, at every voltage a transient simulation over a duration of 
15~ns in 2D cylindrical coordinates has been performed. To calculate from this the CCE 
the simulated transient was corrected for leakage current and integrated over the
above mentioned time interval.

For the final optimisation the I-V and C-V at 455~Hz and 1~kHz and the CCE for the
fluences 3, 6 and $13\cdot 10^{15}$~n$_{eq}$/cm$^2$ at $-20^\circ$C were simultaneously used
and the expression
\begin{equation}
F = \sum_{i,j} w_{i}^j \int_{V_{min}}^{V_{max}}\left(1 -\frac{Q_{i,sim}^j}{Q_{i,meas}^j}\right)^2 \,dV ,
\end{equation}
where $i$ runs over the different fluences and $j$ over the different measurements where 
$Q_{i,sim}^j$ and $Q_{i,meas}^j$ are the simulated and measured quantities, respectively, 
$V_{min}$ the minimum and $V_{max}$ maximum voltage and $w_{i}^j$ weighting factors was 
minimised. The weighting factors can be used to weight the different kind of measurements, but in our case they were usually set to one.

\section{Results}

The final result (see table~\ref{tab:HPTM}) was achieved with an ansatz of 5 traps, 
where the energy levels and types where taken from microscopic measurements and were 
fixed for the optimisation. The free parameter for a trap $k$ 
are the introduction rate g$_{int}^k$, so that the trap concentration is given by 
N$^k$ = g$_{int}^k\Phi_{eq}$, and  the cross sections $\sigma_e^k$, $\sigma_h^k$.
\begin{table}[htb]
  \centering
  \caption{Hamburg Penta Trap Model (HPTM) parameter}
   \begin{tabular}{@{}cccccc@{}}
   \toprule
   Defect & Type & Energy & g$_{int}$ & $\sigma_e$ & $\sigma_h$\\
    &  & & [cm$^{-1}$] & [cm$^{2}$] & [cm$^{2}$] \\
   \midrule
    E30K & Donor & E$_C$-0.1 eV & 0.0497 & 2.300E-14 & 2.920E-16\\
    V$_3$ & Acceptor & E$_C$-0.458 eV & 0.6447 & 2.551E-14 & 1.511E-13\\
    I$_p$ & Acceptor & E$_C$-0.545 eV & 0.4335 & 4.478E-15 & 6.709E-15\\
    H220 & Donor & E$_V$+0.48 eV & 0.5978 & 4.166E-15 & 1.965E-16\\
    C$_i$O$_i$ & Donor & E$_V$+0.36 eV & 0.3780 &3.230E-17 & 2.036E-14\\            
   \bottomrule
  \end{tabular}
  \label{tab:HPTM}
\end{table}
In all simulations the impact ionisation model from van Overstraeten - de Man \cite{vanOverstraeten:1970vq}
 is switched on and for the I$_p$ defect, which is located close to midgap, the trap-assisted tunneling model from Hurkx \cite{Hurkx:1992} with a tunnel mass of 0.25~m$_e$ is used.

\begin{figure}[htb]
  \centering
  \subfloat[]{\label{fig:I_V1}
  \includegraphics[width=0.66\linewidth]{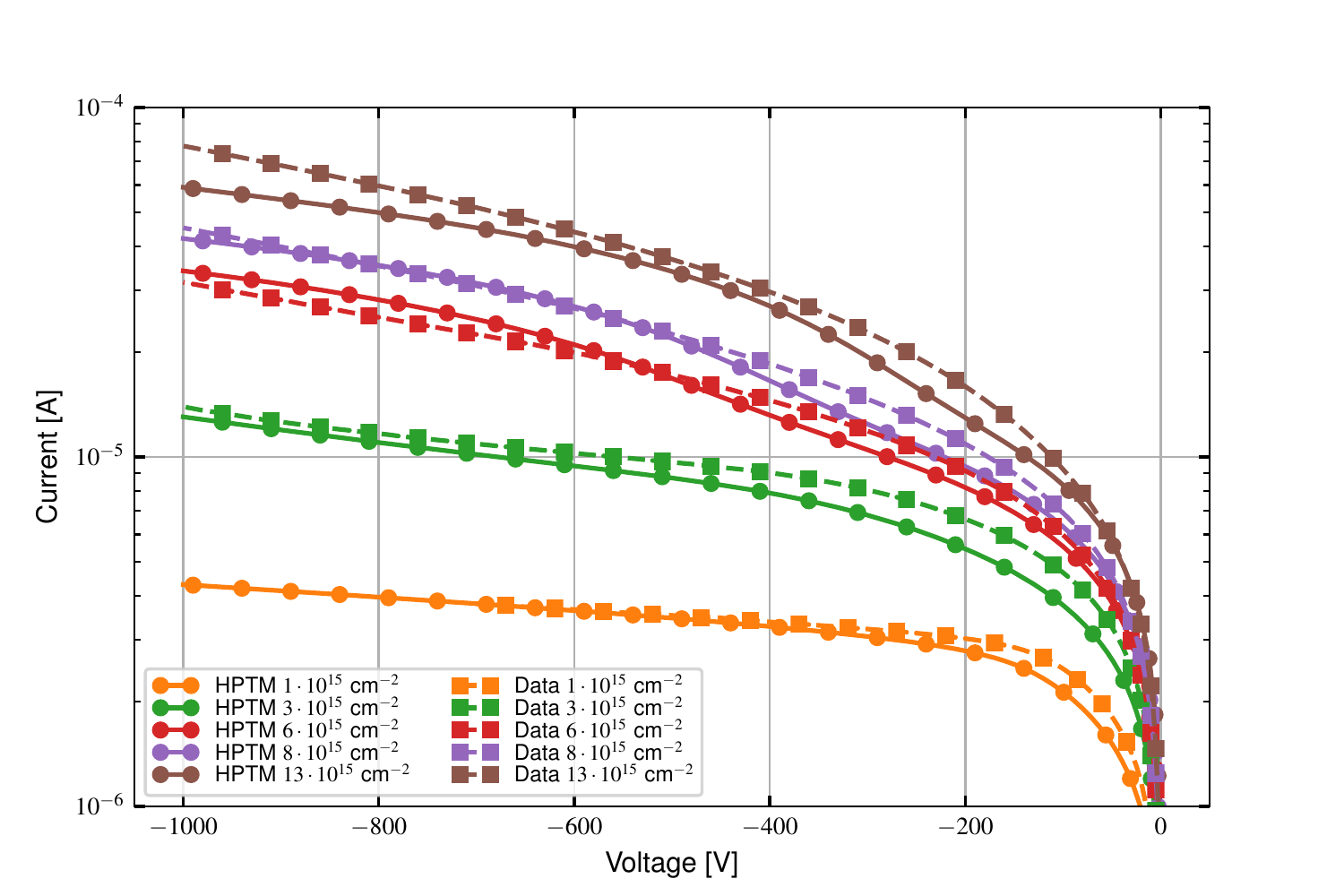}}\\
  \subfloat[]{\label{fig:C_V1}
  \includegraphics[width=0.66
  \linewidth]{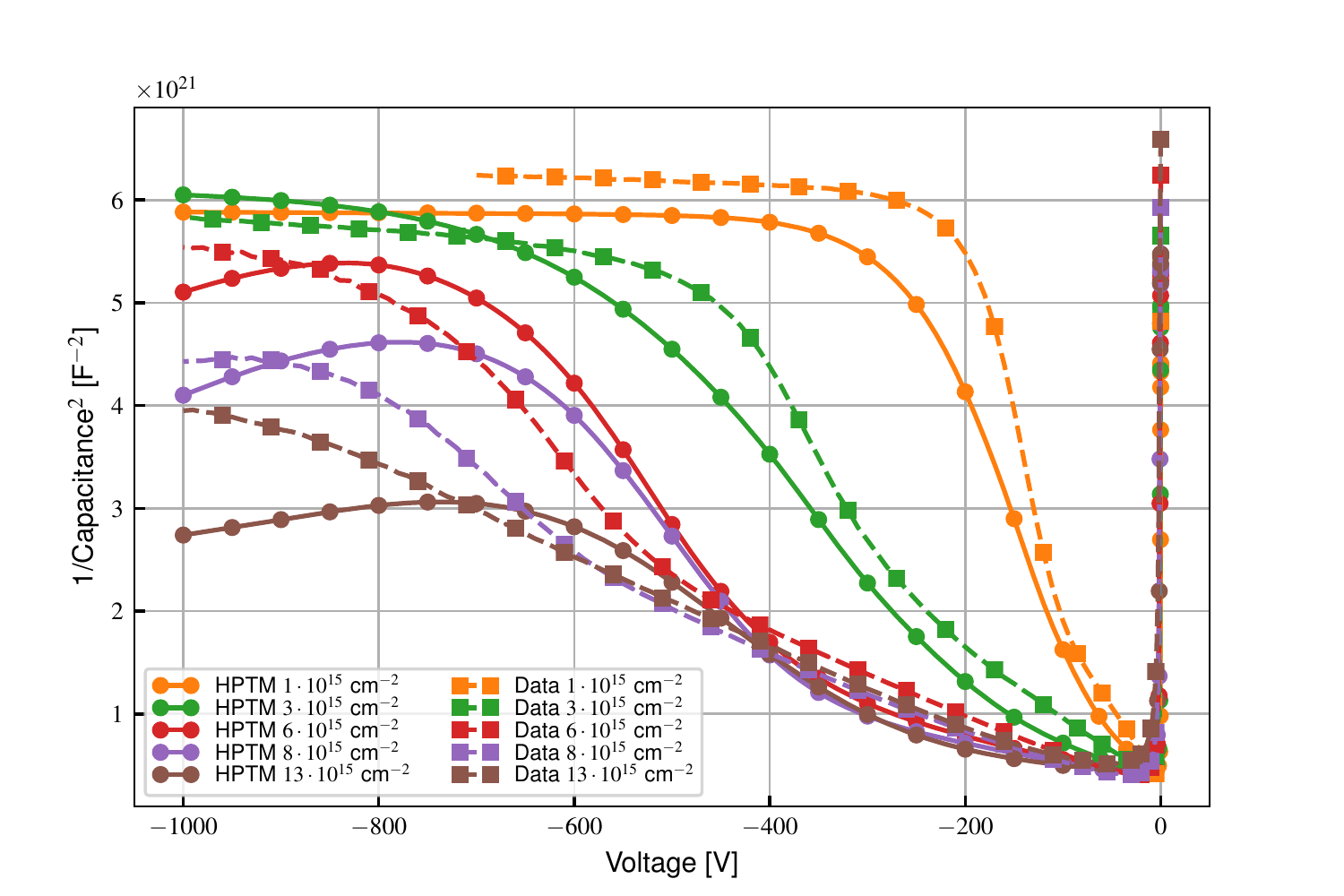}}
  \caption{Measured and simulated I-V (a) and C-V (b) at $-20^\circ$C using the HPTM parameters for p-type diodes irradiated with 24 GeV/c protons to fluences from 1 to $13\cdot 10^{15}$~n$_{eq}$/cm$^2$.}
  \label{fig:IC-V}
\end{figure}

In Fig.~\ref{fig:IC-V}(a) a comparison of the I-V simulated with the Hamburg Penta Trap 
Model (HPTM) parameters and the 
measurements at $-20^\circ$C for fluences from 1 to $13\cdot 10^{15}$~n$_{eq}$/cm$^2$ 
is shown. 
As can be seen, the simulations reproduce the I-V curves approximately over this fluence 
range. 
Similar, in Fig.~\ref{fig:IC-V}(b) the C-V curves at 455~Hz are shown. 
For lower voltages the simulation matches the measurements quite well over the
full fluence range whereas for the higher voltages and fluences a small 
deviation is observed. 
One possible explanation is the onset of impact ionisation at the high voltages.   

\begin{figure}[htb]
  \centering
  \includegraphics[width=0.66\linewidth]{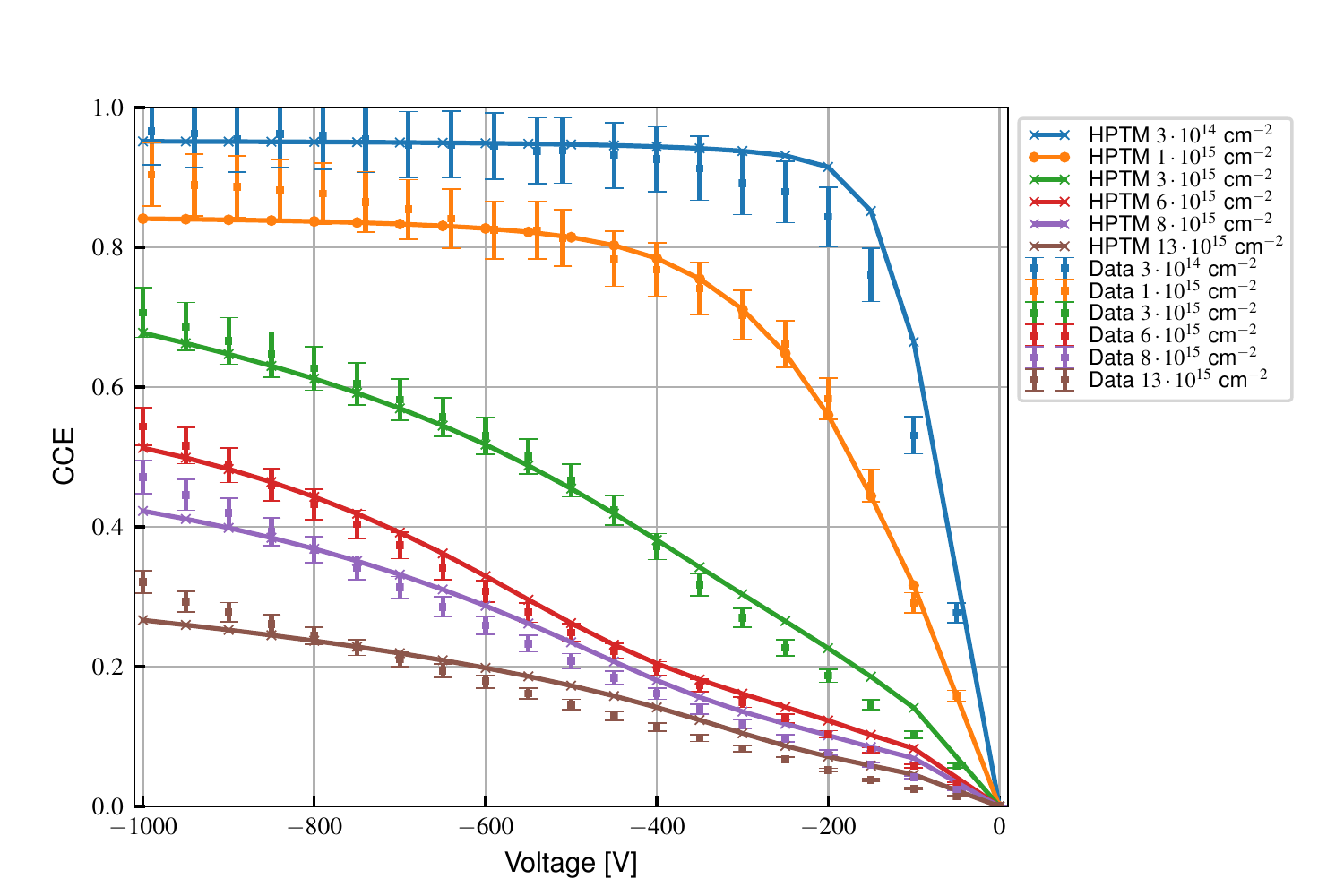}
  \caption{Simulations and measurement of the CCE vs.\ voltage for signals produced by 
  light from an IR laser at $-20^\circ$C using the HPTM parameters for p-type diodes 
  irradiated with 24 GeV/c protons to fluences from 0.3 to $13\cdot 10^{15}$~n$_{eq}$/
  cm$^2$.}
  \label{fig:CCE}
\end{figure}

\begin{figure}[htb]
  \centering
  \includegraphics[width=0.66\linewidth]{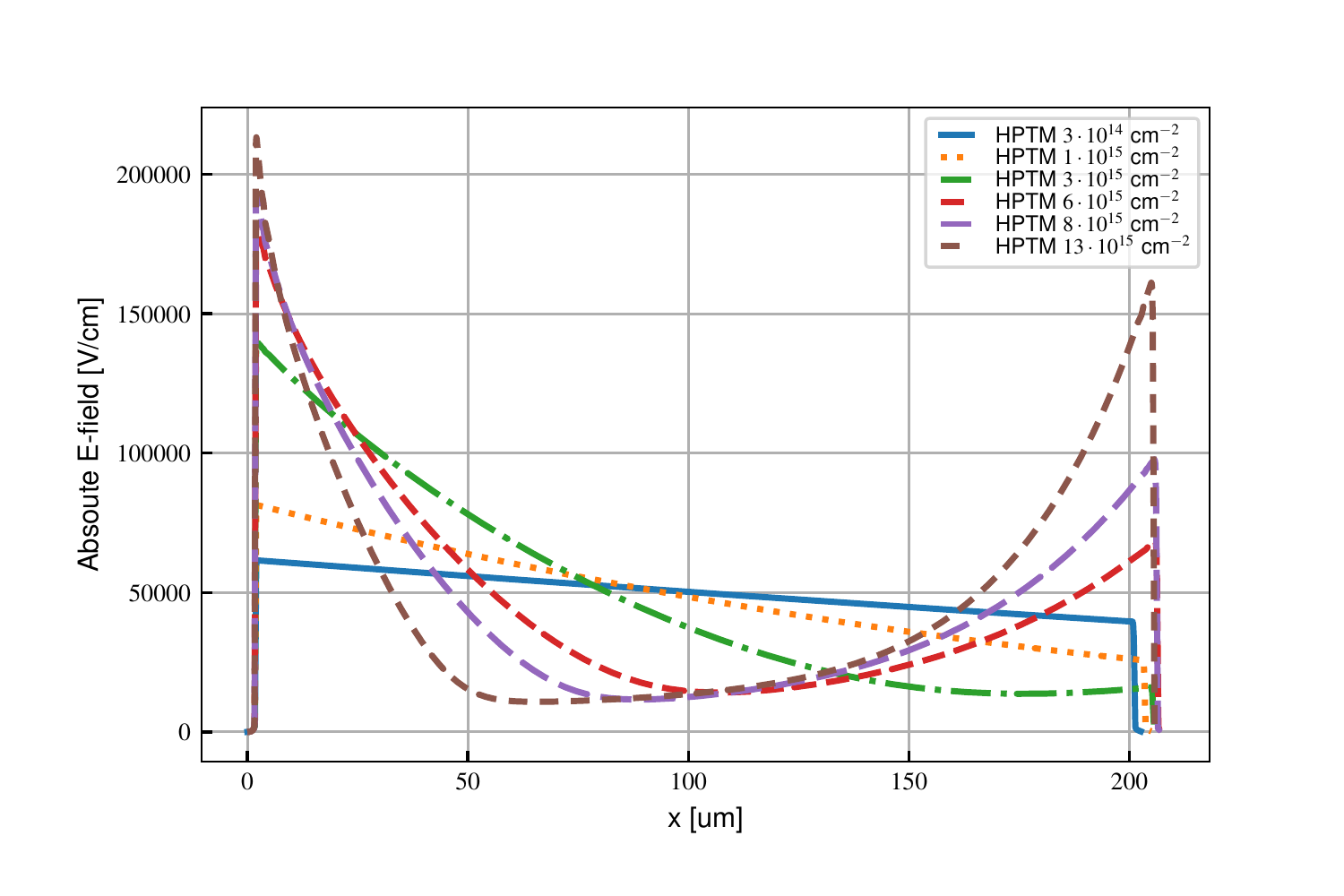}
  \caption{The simulation of the absolute value of the electrical field at $-1000$~V using 
  the HPTM parameters for p-type diodes irradiated with 24 GeV/c protons to fluences from 
  0.3 to $13\cdot 10^{15}$~n$_{eq}$/cm$^2$ at $-20^\circ$C.}
  \label{fig:Efield}
\end{figure}

A correct model must describe the voltage dependence of the CCE and
reproduce the well know double peak structure of the electrical field \cite{Eremin:2002}.
In Fig.~\ref{fig:CCE} the simulations of the CCE for measured signals
produced by light from an IR laser at $-20^\circ$C for the fluences from 0.3 to $13\cdot 
10^{15}$~n$_{eq}$/cm$^2$ are compared to the data. The error bars indicate a
5\% systematic uncertainty. A possible dependence of the absorption length on fluence
has not been taken into account \cite{Scharf:2018}. 
As can be seen the simulations reproduce the voltage dependence qualitatively well 
with the tendency of a slightly too low CCE at high voltages.    
In Fig.~\ref{fig:Efield} the absolute value of the electrical field at -1000~V is 
shown for the fluences from 0.3 to $13\cdot 10^{15}$~n$_{eq}$/cm$^2$. 
The development of the double peak structure with increasing fluence is clearly visible 
and the peak field of the order of 200~kV/cm requires the usage of impact ionisation in 
the simulation.

\begin{figure}[htb]
  \centering
  \includegraphics[width=0.66\linewidth]{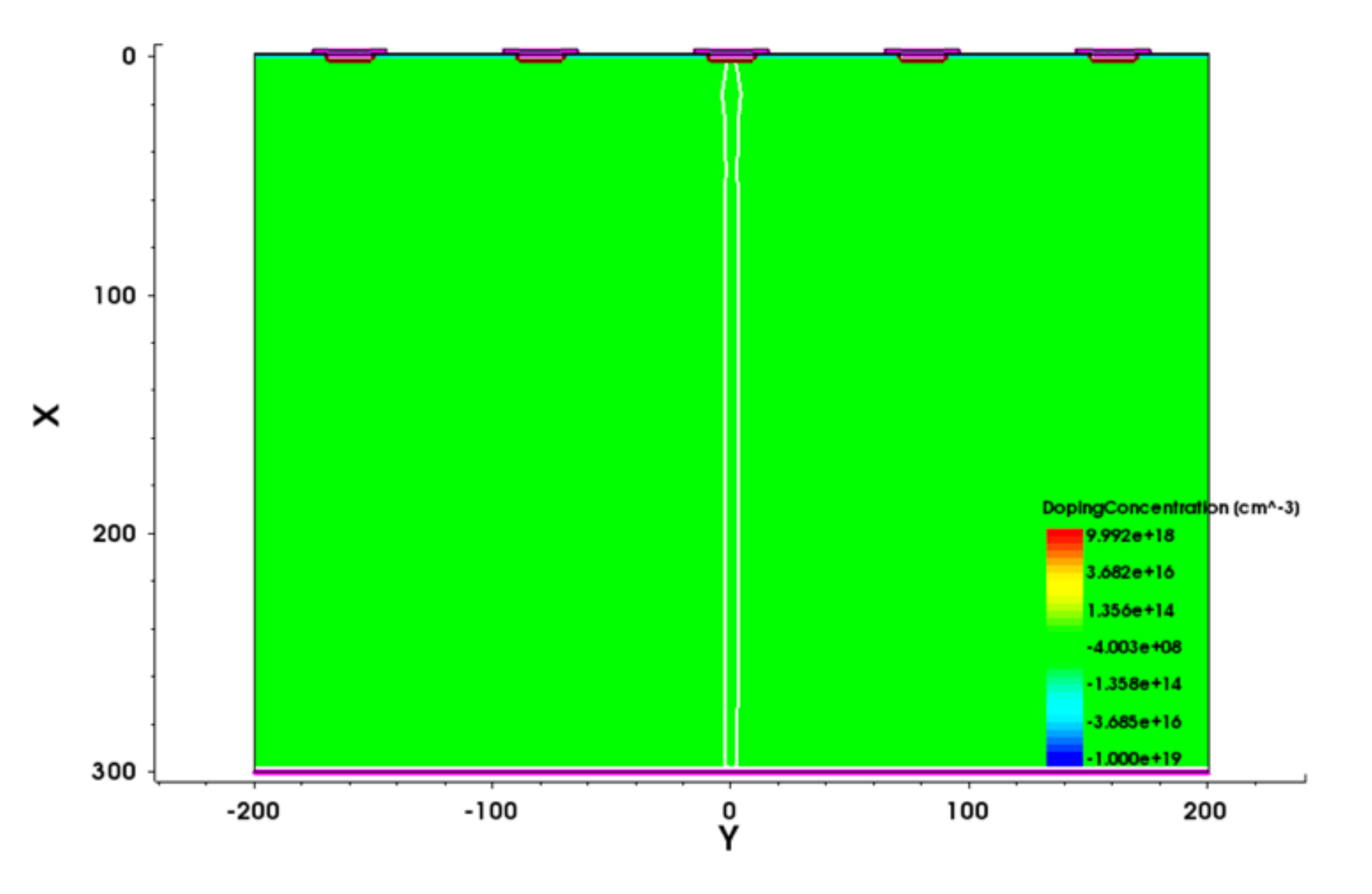}
  \caption{Simulated 2D structure for the validation of the HPTM model with charge 	
  collection measurements on AC-coupled strip sensors. The sensor is 
  300~$\mu$m thick with a pitch of 80~$\mu$m and a n$^+$-implant width of 18~$\mu$m.}
  \label{fig:StripSensor}
\end{figure}

To further validate the HPTM model a comparison of simulated and literature values of 
measured collected charges~\cite{Affolder:2010} of n$^+$-p strip sensors after proton 
irradiation up to a fluence of $2.2\cdot 10^{16}$ n$_{eq}$/cm$^2$ have been performed. 
The charge collection measurements used 300~$\mu$m thick AC-coupled strip sensors with
a pitch of 80~$\mu$m and electrons from a $^{90}$Sr source. 
Measurements were done at a temperature of 248~K with an applied reverse bias of 900~V. 
The strip sensors shown in Fig.~\ref{fig:StripSensor} with a bulk doping of $3\cdot 10^{12}$~cm$^{-3}$ and 5~strips with a pitch of 80~$\mu$m and an n$^+$-implant width 
of 18~$\mu$m were simulated.
The AC-coupling was realised with a 250~nm thick SiO$_2$ in combination with
a 50~nm thick Si$_3$N$_4$. 
To evaluate the CCE a Minium Ionizing Particle (MIP) hitting the center 
of the third strip was simulated using the heavy ion model of Synopsys TCAD and 
adjusting the generated charge to a value of 80~eh-pairs/$\mu$m.

\begin{figure}[htb]
  \centering
  \includegraphics[width=0.66\linewidth]{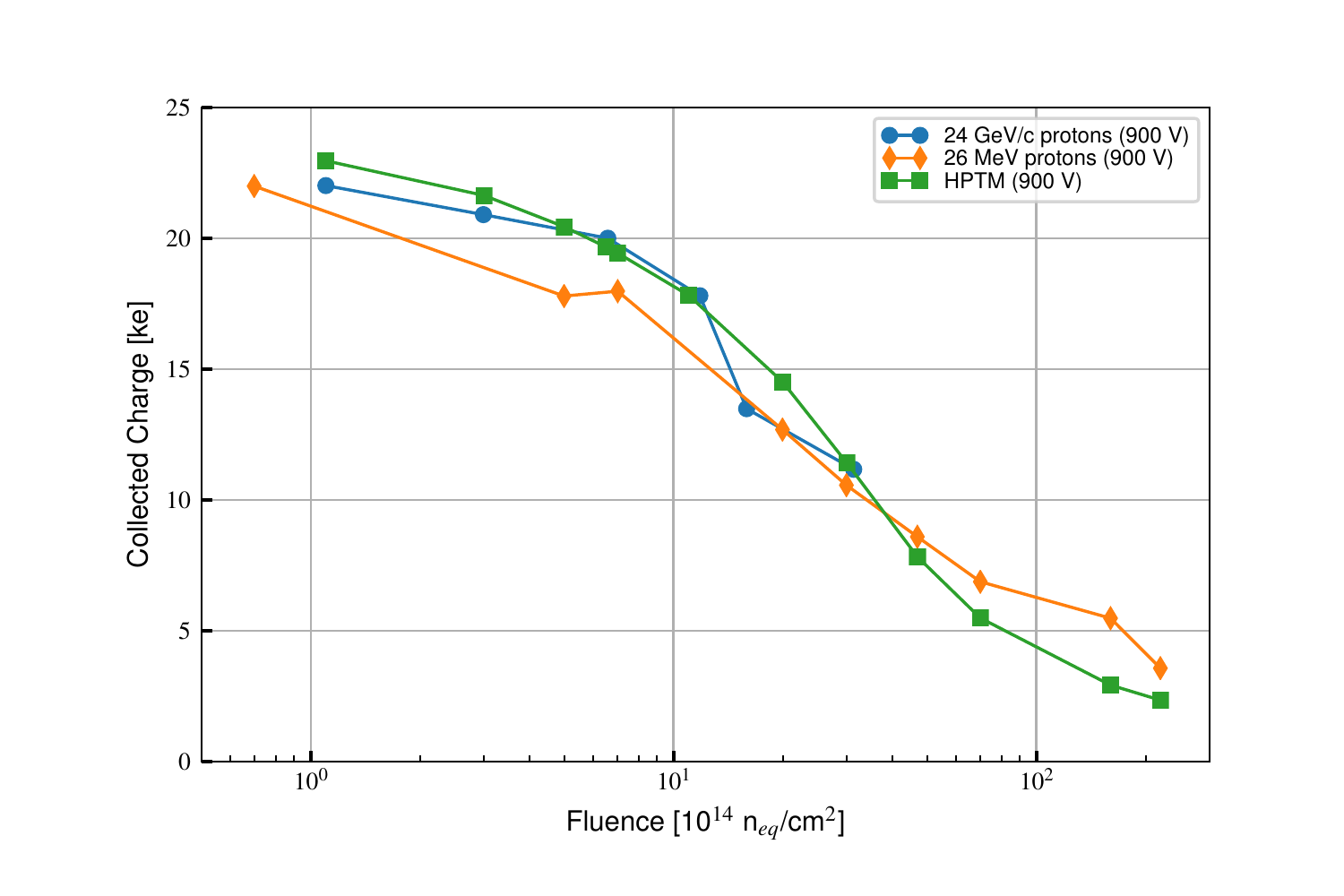}
  \caption{Measured collected charge and simulated with the HPTM model 
  at a bias voltage of 900~V as function of fluence for n$^+$-p strip sensors irradiated 
  with protons.}
  \label{fig:ChargeStrips}
\end{figure}

Figure~\ref{fig:ChargeStrips} shows the measured collected charge and the collected charge 
simulated with the HPTM model at a bias voltage of 900~V as function of fluence for n$^+$-
p strip sensors irradiated with protons of energies of 26~MeV and 23~GeV. As can be seen a 
good agreement over the full fluence range is achieved using the HPTM model.

\section{Conclusion}
In this contribution a radiation damage model with five radiation induced defect states 
is introduced which gives a significantly better and consistent description of a 
large set of measurements of pad diodes irradiated with protons in the fluence range from 
3$\cdot$10$^{14}$~n$_{eq}$/cm$^2$ to 1.3$\cdot$10$^{16}$~n$_{eq}$/cm$^2$ than previous models \cite{Chiochia:2006,Petasecca:2006,Pennicard:2008,
Peltola:2015,Perugia:2015}.
In addition, the application of the HPTM model in the simulation of AC-coupled n$^+$-p strip sensors to model the fluence dependence of proton irradiation gives a good agreement between measurements and simulations. 





\ifCLASSOPTIONcaptionsoff
  \newpage
\fi

\end{document}